\documentclass[preprintnumbers,amsmath,amssymbm,preprint]{revtex4}
\usepackage{epsfig}
\usepackage{graphicx}
\usepackage{amssymb}

\begin{document}

\title{On the status of the hoop conjecture in charged curved spacetimes}
\author{Shahar Hod}
\affiliation{The Ruppin Academic Center, Emeq Hefer 40250, Israel}
\affiliation{ } \affiliation{The Hadassah Institute, Jerusalem
91010, Israel}
\date{\today}

\begin{abstract}
\ \ \ The status and regime of validity of the famous Thorne hoop
conjecture in spatially regular {\it charged} curved spacetimes are
clarified.
\end{abstract}
\bigskip
\maketitle

\section{Introduction}

The hoop conjecture \cite{Thorne} has attracted the attention of
physicists and mathematicians since its introduction by Thorne
almost five decades ago \cite{Thorne,Mis}. This mathematically
elegant and physically influential conjecture asserts that a
self-gravitating matter configuration of mass $M$ will form an
engulfing horizon if its circumference radius $R=C/2\pi$ is equal to
(or less than) the corresponding Schwarzschild radius $2M$
\cite{Noteunit}. That is, the hoop conjecture states that
\cite{Thorne}
\begin{equation}\label{Eq1}
C\leq 4\pi M\ \  \Longrightarrow \ \ \text{Black-hole horizon
exists}\ .
\end{equation}

It is widely believed that the hoop conjecture reflects a
fundamental aspect of classical general relativity. In particular,
the conjecture is supported by several studies (see
\cite{Red,Abr,Hod1} and references therein). Intriguingly, however,
there are also some claims in the physics literature that the Thorne
hoop conjecture can be violated in {\it charged} curved spacetimes
\cite{Leon,Bon}.

The main goal of the present compact paper is to clarify the status
of the Thorne hoop conjecture in spatially regular charged
spacetimes. In particular, below we shall explicitly demonstrate
that the hoop conjecture is valid in charged curved spacetimes
provided the mass parameter on the r.h.s of the hoop relation
(\ref{Eq1}) is appropriately interpreted as the gravitational mass
$M(R)$ contained within the engulfing hoop of radius $R$ and {\it
not} as the total (asymptotically measured) mass $M_{\infty}$ of the
entire spacetime.

\section{Validity of the hoop conjecture in spatially
regular charged spacetimes}

It has been argued in \cite{Leon,Bon} that the hoop conjecture
(\ref{Eq1}) can be violated in spatially regular horizonless charged
spacetimes. In particular, Ref. \cite{Leon} analyzed the compactness
of spherically symmetric fluid matter configurations with uniform
charge densities and concluded that, taking the mass parameter in
the r.h.s of the hoop relation (\ref{Eq1}) as the {\it total} mass
$M_{\infty}$ of the system (as measured by asymptotic observers),
the hoop conjecture can be violated. As an illustrative example,
Ref. \cite{Leon} constructed a uniformly charged horizonless ball
which is characterized by the dimensionless relations
\begin{equation}\label{Eq2}
{{M_{\infty}}\over{R}}=0.65\ \ \ \ \text{and}\ \ \ \
{{Q^2}\over{R^2}}=0.39\  .
\end{equation}
This spatially regular {\it horizonless} charged matter
configurations is characterized by the dimensionless ratio
\begin{equation}\label{Eq3}
{{C}\over{4\pi M_{\infty}}}\simeq 0.769<1\  ,
\end{equation}
and, as claimed in \cite{Leon}, it therefore violates the hoop
conjecture (\ref{Eq1}).

However, we believe that in the Thorne hoop conjecture (\ref{Eq1}),
which relates the mass parameter of the system to its circumference
radius $R=C/2\pi$, it is physically more appropriate to interpreted
$M$ as the gravitational mass contained {\it within} the engulfing
hoop and not as the total mass of the entire curved spacetime.

In particular, since the exterior $(r>R)$ electromagnetic energy
density associated with a charged ball of radius $R$ and electric
charge $Q$ is $T^0_0(r>R)=Q^2/8\pi r^4$ \cite{Got}, the
electromagnetic energy $E_{\text{elec}}(r>R)=\int^{\infty}_{R}T^0_0
4\pi r^2dr$ {\it outside} the charged ball is given by the simple
expression
\begin{equation}\label{Eq4}
E_{\text{elec}}(r>R)={{Q^2}\over{2R}}\  .
\end{equation}
Thus, for a charged ball of radius $R$, electric charge $Q$, and
{\it total} mass (energy) $M_{\infty}$ as measured by asymptotic
observers, the gravitational mass contained {\it within} $(r\leq R)$
the ball is given by
\begin{equation}\label{Eq5}
M(r\leq R)=M_{\infty}-{{Q^2}\over{2R}}\  .
\end{equation}

From Eqs. (\ref{Eq2}) and (\ref{Eq5}) one obtains the dimensionless
relation
\begin{equation}\label{Eq6}
{{M(r\leq R)}\over{R}}=0.455\
\end{equation}
for the horizonless charged matter configuration considered in
\cite{Leon}. Taking cognizance of Eqs. (\ref{Eq1}) and (\ref{Eq6}),
one finds
\begin{equation}\label{Eq7}
{{C(R)}\over{4\pi M(r\leq R)}}\simeq 1.099>1\  .
\end{equation}
The dimensionless ratio (\ref{Eq7}) implies, in particular, that the
uniformly charged matter configurations studied in \cite{Leon} do
{\it not} violate the Thorne hoop conjecture (\ref{Eq1}).

\section{Summary}

In this compact paper we have explored the (in)validity of the
Thorne hoop conjecture \cite{Thorne} in spatially regular {\it
charged} curved spacetimes. Our analysis is motivated by the
intriguing claims made in the physics literature (see e.g.
\cite{Leon,Bon}) according to which this famous conjecture, which is
widely believed to reflect a fundamental aspect of classical general
relativity, can be violated by horizonless charged matter
configurations.

The present analysis clearly demonstrates the fact that, as opposed
to the claims made in \cite{Leon,Bon}, the Thorne hoop conjecture is
valid in charged spacetimes provided that, for a given radius $R$ of
the engulfing hoop, the mass parameter in the hoop relation
(\ref{Eq1}) is appropriately interpreted as the gravitational mass
$M(R)$ contained {\it within} the hoop (sphere) of radius $R$ and
not as the total mass $M_{\infty}$ of the entire spacetime.

\bigskip
\noindent {\bf ACKNOWLEDGMENTS}

This research is supported by the Carmel Science Foundation. I would
like to thank Yael Oren, Arbel M. Ongo, Ayelet B. Lata, and Alona B.
Tea for stimulating discussions.


\begin{thebibliography}{99}
\bibitem{Noteunit} We shall use natural units in which $G=c=\hbar=1$.
\end{thebibliography}
\end{document}